\documentclass[aps,pre,twocolumn,showpacs,superscriptaddress,groupedaddress,reprint]{revtex4-1}
\usepackage{graphicx}
\usepackage{mathrsfs}
\usepackage{natbib}
\usepackage{color}
\usepackage{epstopdf}
\usepackage{cleveref}
\usepackage{amsmath}
\usepackage{epsfig}

\begin{document}

\widetext

\title {Stability and anomalous entropic elasticity of sub isostatic random-bond
  networks}

\author{M. C. Wigbers$^{1}$} 
\author{F.C. MacKintosh$^{1}$} 
\author{M. Dennison$^{1}$$^{2}$} 

\affiliation{$^{1}$Department of Physics and Astronomy,
  VU University, De Boelelaan 1081, 1081 HV Amsterdam, The
  Netherlands}
\affiliation{$^{2}$Department of Applied Physics and Institute for Complex Molecular Systems, Eindhoven University of Technology,
  P.O. Box 513, NL-5600 MB Eindhoven, The Netherlands}

\date{\today}

\begin{abstract}
  We study the elasticity of thermalized spring networks under an
  applied bulk strain. The networks considered are sub-isostatic
  random-bond networks that, in the athermal limit, are known to have
  vanishing bulk and linear shear moduli at zero bulk strain. Above a
  bulk strain threshold, however, these networks become rigid,
  although surprisingly the shear modulus remains zero until a second,
  higher, strain threshold. We find that thermal fluctuations
  stabilize all networks below the rigidity transition, resulting in
  systems with both finite bulk and shear moduli. Our results show
  a $T^{0.66}$ temperature dependence of the moduli in the
  region below the bulk strain threshold,
  resulting in networks with anomalously high rigidity as
  compared to ordinary entropic elasticity. Furthermore we find a
  second regime of anomalous temperature scaling for the shear modulus
  at its zero-temperature rigidity point, where it scales as
  $T^{0.5}$, behavior that is absent for the bulk modulus since its
  athermal rigidity transition is discontinuous.
\end{abstract}

\maketitle

\section{Introduction}
Materials such as plastics and rubbers as well as tissues and living
cells contain polymer networks, which, among other roles, provide
structural support to these materials. Tissues and cellular networks
are especially sensitive to external stresses
\cite{ref:Fung1967,ref:Gardel2004,ref:Storm2005,ref:Gardel2006,ref:Broedersz2008,ref:Kasza2009,ref:Lin2010}, and a number of
theoretical and simulation studies have attempted to gain an
understanding of what controls the response of such systems to
deformations \cite{ref:Kroy1,ref:ChaseReview}. In 1864, Maxwell showed
that there is a connectivity threshold $z_{c}$, determined by the
average coordination number of the network nodes, at which athermal
networks of springs become rigid \cite{ref:Maxwell}. This threshold,
referred to as the isostatic point, occurs when the number of degrees
of freedom of the network nodes are just balanced by the number of
constraints arising from the springs. This purely mechanical argument
has been used to describe the stability systems ranging from emulsions
and jammed particle packings \cite{ref:Hecke,ref:Liu} to amorphous
solids \cite{ref:Lubensky_as} and folded proteins
\cite{ref:Rader_pf}. Beyond this, theoretical work has shown that
there are numerous ways of stabilizing a network, and therefore tuning
its rigidity, below the isostatic point \cite{ref:Wyart}. Examples include the addition
of a bending stiffness to the model filaments
\cite{ref:Fred_mik1,ref:Frey_mik1,ref:Chase_bend}, by applying
 stress \cite{ref:Alexander}, either internally \emph{via} molecular motors
\cite{ref:Chase_motor,ref:Misha_motor} or externally by placing the network under
tension by applying a bulk strain to the system
\cite{ref:Misha_bulk}. It has been shown that a network's rigidity
point can be shifted from the Maxwell point by adding these
interactions and forces to the system. In the case of applying a bulk
strain \cite{ref:Misha_bulk} the system can be stabilized by
stretching the network until all the floppy modes have been pulled
out, resulting in a critical strain at which the network is just
rigid.

In addition to these athermal models, recent work has shown how
temperature can stabilize a mechanically floppy network
\cite{ref:Plischke,ref:Farago2,ref:Farago3,ref:Farago4,ref:PRL}. In
Ref. \cite{ref:PRL} it was found that at and below the isostatic point
the network response to deformation, defined by the shear modulus, not
only becomes finite when thermal fluctuations are present, but that it
also shows an anomalous temperature scaling of $T^{\alpha}$, where
$\alpha<1$. This sub-linear temperature dependence indicates that a
network would exhibit a \emph{larger} resistance to deformation than
would be expected from entropic elasticity, where one would expect a
linear temperature dependence \cite{ref:deGennes}. The origin of this
anomalous temperature dependence remains unclear, and in addition
there have been few studies into the effects of thermal fluctuations
on sub-isostatic networks
\cite{ref:Rubinstein,ref:Barriere,ref:Plischke,ref:Tessier}. Furthermore,
in Ref.~\cite{ref:PRL} a triangular lattice based network was used,
and an open question is how general the anomalous regimes found are,
since network architecture can have vast effects on a systems response
to deformation \cite{ref:Heussinger1,ref:Heussinger2}.

In this paper, we study the effects of thermal fluctuations on an
under-constrained and mechanically floppy random-bond network. The
architecture of a random-bond network is as different as possible from
a triangular lattice network, as the nodes are arranged isotropically
and their is a distribution of filament lengths. The random-bond model
proposed by Jacobs and Thorpe \cite{ref:Thorpe_RB} has been used previously to study the
effects of applying a bulk strain on the rigidity of athermal networks
\cite{ref:Misha_bulk}. The connectivity threshold for rigidity perculation of
  this model will be somewhat lower than that of a lattice network
  \cite{ref:Rivoire,ref:Kasiviswanathan}.


We study the bulk strain and temperature dependence of the internal pressure, bulk modulus and shear modulus of random-bond networks with average coordination number of z = 3. This coordination number lies between the connectivity percolation threshold (below which the networks would be floppy regardless of strain or thermal fluctuations \cite{ref:Alexander,ref:Misha_bulk}) and the isostatic threshold for central force interactions (above which athermal spring networks become rigid\cite{ref:Maxwell}).

\begin{figure}[h!]
\includegraphics[width=0.5\textwidth]{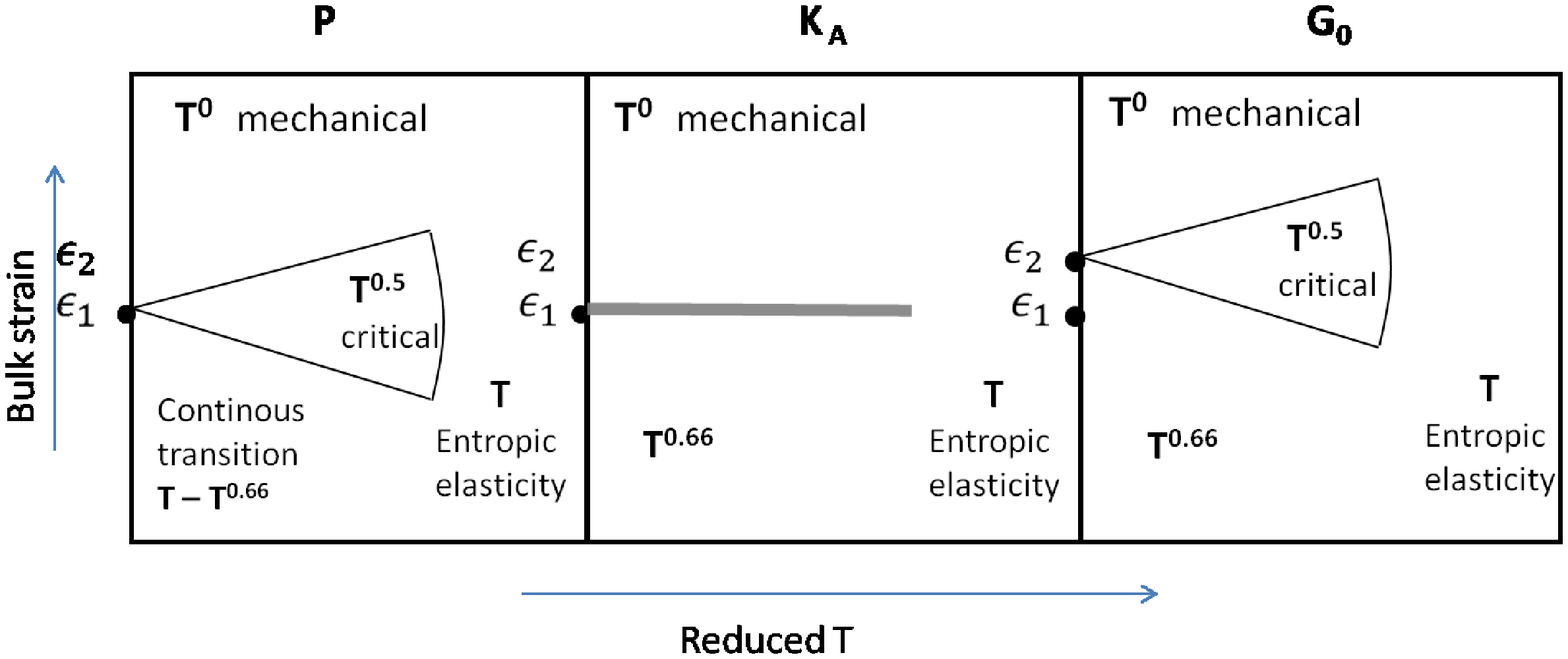}
\caption{Schematic phase diagram showing behavior of the internal
  pressure $P$, bulk modulus $K_{A}$ and shear modulus $G$ with bulk
  strain $\epsilon$ and temperature $T$ for sub-isostatic random-bond
  networks with connectivity $z=3$.}
\label{fig:phase diagram}
\end{figure}

We show that, as reported previously in
Ref.~\cite{ref:Misha_bulk}, there exists a bulk strain threshold at
which the system will begin to resist bulk deformations at zero
temperature. However, the network does not begin to resist shear
deformation until a second, higher, strain threshold is reached, and
it is these two thresholds that control the network response to the
applied deformations. We find anomalous scaling regimes for the shear
modulus at and below its threshold, similar to the results of
Ref.~\cite{ref:PRL}, where the bulk strain applied to the networks in
this study takes on a similar role to the connectivity in
Ref.~\cite{ref:PRL}. Interestingly, we find that, while the bulk
modulus exhibits a similar anomalous scaling regime below its
threshold, we find no temperature dependence at its strain threshold,
at which there is a first order zero-temperature rigidity
transition. The network behavior is summarized in the phase diagrams
shown in Fig.~\ref{fig:phase diagram}.

\section{Physical Picture}
Since Maxwell \cite{ref:Maxwell} it has been known that an athermal
network of central-force springs will be floppy below a critical,
isostatic connectivity threshold. This means that there is no energy
cost for small bulk or shear deformations. When applying an
increasingly large uniform bulk strain, such networks will begin to
resist additional bulk deformations at a strain threshold
corresponding to a rigidity transition \cite{ref:Misha_bulk}, at which
the network will be just rigid. Applying small deformations on a rigid
network will cost energy, since the springs will be stretched, which
results in a stable network exhibiting a non-zero bulk modulus at zero
temperature.

A mechanically floppy network will also be stabilized by thermal
fluctuations \cite{ref:PRL,ref:Plischke}. The resulting network is
rigid both above \emph{and} below the rigidity point. A deformation of
a mechanically floppy network results in a reduction of the number of
micro states that the system can assume, even though the system energy
remains unchanged. This results in a change in entropy as the system
is deformed, which gives rise to a change in the free energy,
resulting in non-zero elastic moduli at finite temperatures. Thus,
below the rigidity point the network is stabilized by thermal
fluctuations, as the entropic contribution to the moduli dominate over
the mechanical contribution. When the network is sufficiently
stretched, i.e., above the bulk strain rigidity threshold, all springs
are under tension that causes the mechanical stretching energy
(controlled by the spring constant) to dominate the thermal
fluctuations in stabilizing the network, and the network rigidity then
becomes independent of temperature. As thermal networks are always
rigid, there is no bulk strain threshold at which the network becomes
stable.  However, if a network is taken to the rigidity point, we find
that there can be an anomalous intermediate regime in which the
network is stabilized by both temperature and the spring constant.

These three different regimes of network stability are defined by the
bulk strain at the zero temperature rigidity transition. This strain depends on how constrained the system is,
controlled, for example, by varying the connectivity of the network by
changing the number of springs. Lowering the connectivity will lower
the number of constraints in the network and it has been shown that
sub isostatic networks with increasingly lower connectivities need to
be stretched increasingly more to become rigid \cite{ref:Misha_bulk}.

\section{The Model}
\begin{figure}[h!]
\includegraphics[width=0.4\textwidth]{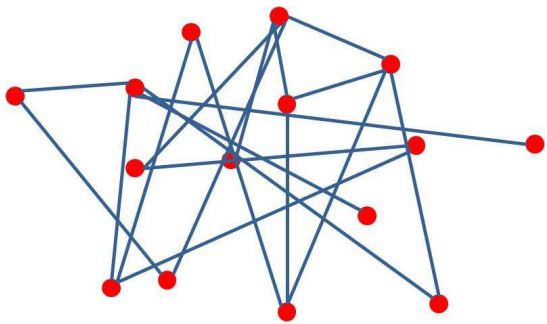}
\caption{(Color online) Schematic representation of a random-bond network. The network
  is constructed by placing $N$ nodes randomly in a box. The nodes are
  randomly connected by $zN/2$ springs to reach an average
  connectivity $z$.}
\label{network}
\end{figure}

In this paper we study the effects of thermal
fluctuations and bulk strain on the stability of sub-isostatic
random-bond networks. The random-bond network is constructed by
placing $N$ nodes randomly in a $2$ dimensional box of area $A$, which
are then connected by $N_{sp}$ springs until the network reaches an
average connectivity $z=2N_{sp}/N$\cite{ref:Thorpe_RB,ref:Misha_bulk}.
Since unconnected nodes will not contribute to the networks response,
each node is first connected to at least one randomly chosen other
node. Thus, our networks are random, in that both the positions of the nodes and the length of the connecting springs are random. Periodic boundary conditions are used throughout, and the
springs may cross the system boundaries. Furthermore, we do not allow
two nodes to be connected by more than one spring, nor that both ends
of the spring connect to the same node. This method would still allow
for disconnected clusters to form.While this method
  does not generate a truly random network, we find that in practice
  this does not effect the results we present in this paper, as will
  be shown.  A schematic of a random-bond network is shown in
Fig.~\ref{network}. The springs have a rest length $l_0$, which will
vary for each spring and, by construction, the average rest length
will be half the system size. We use the average spring length
$\langle l_0\rangle$ as the unit of length, and we note that for
systems with the same density of nodes $\langle l_0\rangle$ grows as
$\sqrt{N}$, and as such there is no well defined thermodynamic
limit. In this simple model the only two energy scales are the
stretching energy and the thermal energy. The total energy of the
network is given by the sum of the energy of all $N_{sp}$ springs

\begin{equation}
U = \frac{k_{\rm sp}}{2}\sum^{N_{sp}}_{i}\langle l_i-l_{0,i}\rangle ^2.
\label{mechenergy}
\end{equation} 
where $k_{sp}$ is the spring constant and $l_{0,i}$ the rest length of
spring $i$ which has length
\begin{equation}
l_{i} =\sqrt{{(x_{2}-x_{1})^2 + (y_{2}-y_{1})^2}}, 
\end{equation}
where $x_{j},y_{j}$ are the coordinates of nodes $j=1,2$ that are
connected by spring $i$. To study fiber networks, it is common to set the spring constant to  $k_{sp,i} = \mu/l_{0,i}$ \cite{ref:Chase_bend,ref:Frey_mik1,ref:Fred_mik1,broedersz2012filament,conti2009cross,ref:Heussinger1,ref:Heussinger2}, where $\mu$ is the 1D Young's (stretch) modulus. This means that long springs will become progressively weaker and contribute less to the network response. For polymers, flexible or semi-flexible, yet different length dependence is possible \cite{ref:Mac_PRL,ref:Gardel2004,ref:HuismanPRE,huisman2011internal}. We chose to keep the spring constant
  identical for each spring, as we find that this has no qualitative
  effect on our results (see Fig. 3) and speeds up our computer simulations.


This network architecture is isotropic and differs qualitatively from
a lattice-based network, for which the springs have either the same
length or a narrow discontinuous distribution of lengths. In the
random-bond network, there are springs with lengths of the order of
the system size, which would prevent network collapse at finite
temperature due to entropic forces \cite{ref:Boal,ref:PRL}. Thus, the
random-bond model is stable to thermal fluctuations without an imposed
tension at the boundaries. We have chosen this
minimalist off-lattice network in order to study the anomalous low
temperature behavior found in lattice networks \cite{ref:PRL}. However,
this network is most different from a lattice network, this network
shows some similar behavior with temperature. It is deliberately
highly theoretical, and does not represent a real system, but does let
us examine the anomalous temperature dependence in detail.

In order to study the entropic stabilization we
apply varying bulk strains to the system by uniformly scaling the
system area such that

\begin{equation}
\label{area}
A = A_{0}(1+\epsilon)^{2},
\end{equation}
where $A_{0}$ is the rest area of a fully relaxed network at
temperature $T=0$, and $\epsilon$ is the applied strain. The $x$- and
$y$- coordinate of each node are also scaled, defining new coordinates
$x^{\prime}$ and $y^{\prime}$ for node $j$ as
\begin{equation}
x^{\prime}_{j} = x_{j} (1+ \epsilon).
\end{equation}
We introduce the temperature $T$ using Monte Carlo simulations to study
the equilibrium behavior of thermal systems.

\subsection{Elastic moduli and internal pressure}
We determine the internal pressure and bulk modulus of the system
under bulk strain. The internal pressure is defined as
\begin{equation}
P=-\frac{\partial F}{\partial A},
\end{equation}
where $F$ is the Helmholtz free energy, and can be calculated in our
simulations as \cite{ref:Press}
\begin{eqnarray}
P = \frac{N}{A} k_B T + \frac{1}{2A} \displaystyle \sum^N_i \displaystyle \sum^N_j \langle f_{i,j} \cdot l_{i,j} \rangle \nonumber \\
=  \frac{N}{A} k_B T - \frac{1}{2A} \displaystyle \sum^{N_{\rm sp}}_k \left \langle k_{\rm sp} l_{k} (l_{k} - l_{0})\right \rangle,
\end{eqnarray}
where the first line contains a sum over all pairs of nodes and
the second line contains a sum over all springs, since the
force $f_{i,j}$ between node $i$ and node $j$ is only non-zero if
there is a spring connecting $i$ and $j$. The first term represents
the ideal gas behavior and the second term corrects for spring
interactions. By calculating the internal pressure at various areas we
can then calculate the bulk modulus, which is defined by
\begin{equation}
\label{eq:bulk}
K_A = -A \frac{\partial P}{\partial A}.
\end{equation}

In addition, we calculate the shear modulus $G$ of the networks at
each bulk strain. $G$ is defined by
\begin{equation}
G = \frac{1}{A}\frac{\partial^2 F}{\partial\gamma^2},
\end{equation}
where $\gamma$ is the shear strain. In order to shear the network we
use Lees-Edwards boundary conditions \cite{ref:Lees_Edwards}, where the
energy of the springs crossing the top boundary of the simulation box
is modified to become
\begin{equation}
E_{\rm sp}(l) = \frac{k_{\rm sp}}{2} \left ( \sqrt{(x_{ij} + \gamma L_y)^2 +
  (y_{ij})^2}- l_0 \right)^2,
\end{equation}
where $L_y$ is the height of the simulation box. We initially shear
the networks at zero temperature, obtaining a configuration under shear,
and then increase the temperature from zero. For these thermal systems, we
calculate the shear stress $\sigma$ as in
Refs.~\cite{ref:G,ref:PRL}. The shear modulus can then be calculated
by taking the derivative of the stress on the network with respect to
$\gamma$ at $\gamma=0$.

\section{Results}

\begin{figure}[h!]
\includegraphics[height=0.5\textwidth,angle=270]{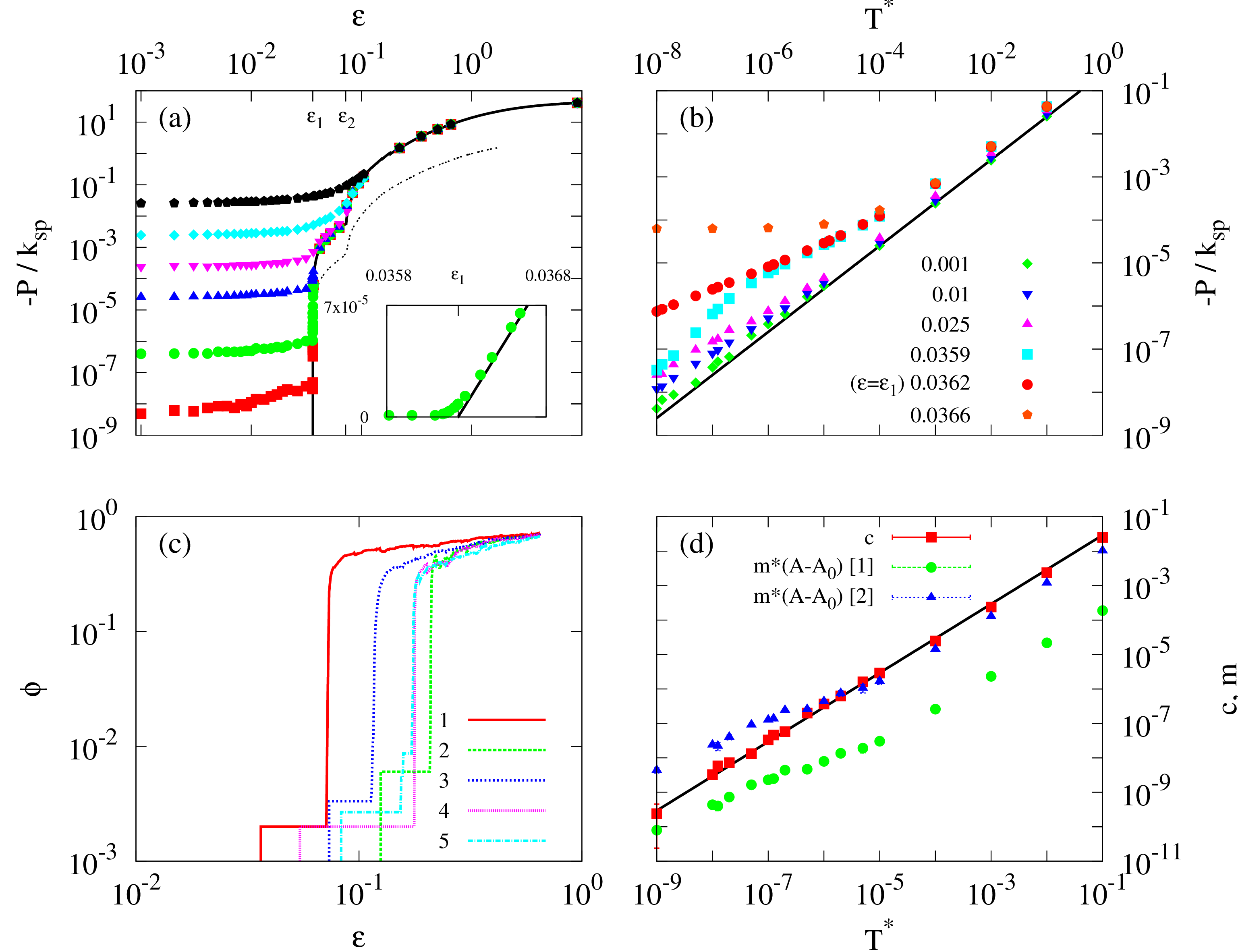}
\caption{(Color online) Pressure $P$ of random-bond spring network with $N=1000$
  nodes against (a) bulk strain $\epsilon$ and (b) reduced temperature
  $T^{*}=k_{B}T/k_{sp}\langle l_0\rangle^2$. Solid line in (a) shows
  zero temperature behavior, points are for thermal systems,
  while the dashed line shows results for the same
    system but where the spring constant of individual springs is
    given by $k_{sp,i}=\mu/l_{0,i}$, where $\mu$ is a stretch
    modulus. Solid line in (b) shows linear $T$ dependence. (c)
  Fraction of stretched springs $\phi$ in network as the a function of
  $\epsilon$. (d) Behavior of $m$ and $c$ in the function $-P =
  m(A-A_{0}) + c$ with reduced temperature $T^{*}$, where $A$ is the
  area and $A_{0}$ is the rest area at $\epsilon=0$. $[1]$ indicates
  $\epsilon=0.001$, $[2]$ indicates $\epsilon=0.03$. Solid line shows
  linear $T$ dependence.}
\label{fig_plotPe}
\end{figure}

\begin{figure}[h!]
\includegraphics[height=0.4\textwidth,angle=270]{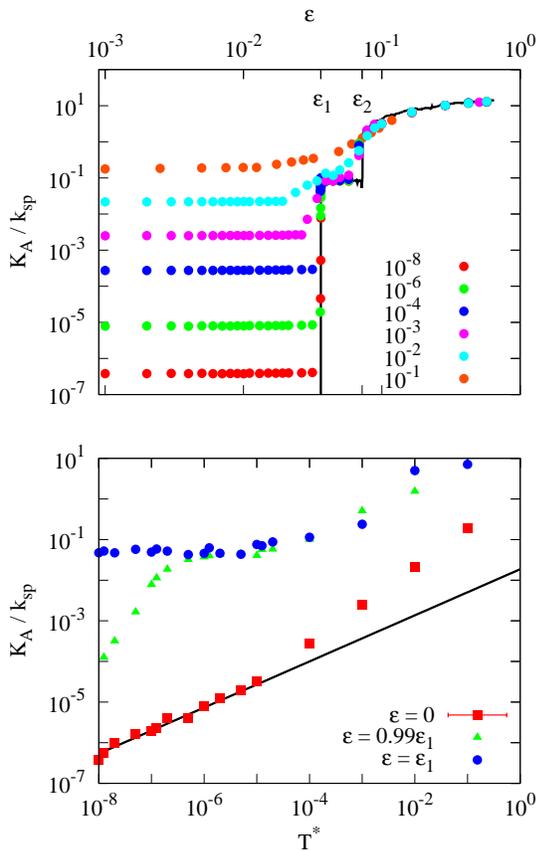}
\caption{(Color online) Bulk modulus $K_{A}$ of random-bond spring network with
  $N=1000$ nodes against (a) bulk strain $\epsilon$ and (b) reduced
  temperature $T^{*}=k_{B}T/k_{sp}\langle l_0\rangle^2$. Solid line in
  (a) shows zero temperature behavior, points are for thermal
  systems. Solid line in (b) shows $T^{0.66}$ depdendence.}
\label{fig_plotKe}
\end{figure}

\begin{figure}[h!]
\includegraphics[height=0.4\textwidth,angle=270]{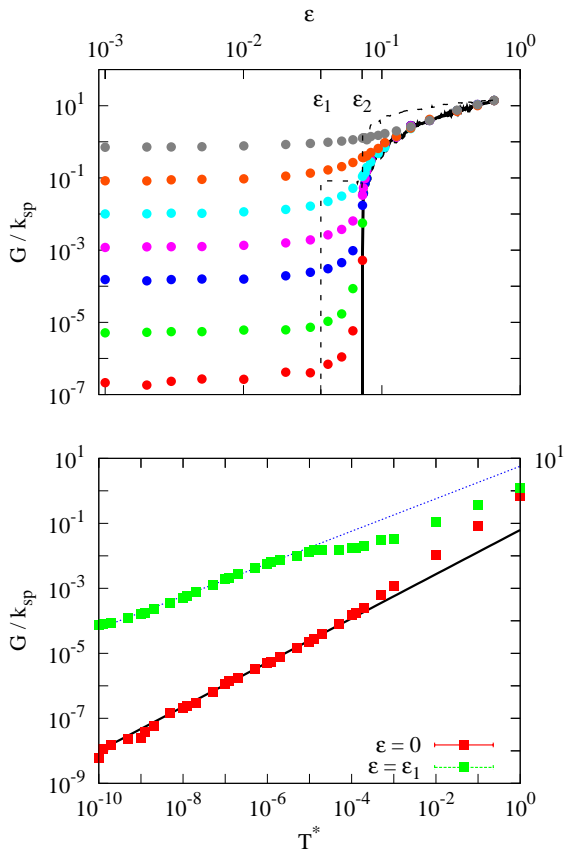}
\caption{(Color online) Shear modulus $G$ of random-bond spring network with $N=1000$
  nodes against (a) bulk strain $\epsilon$ and (b) reduced temperature
  $T^{*}=k_{B}T/k_{\rm sp}\langle l_0\rangle^2$. Solid line in (a)
  shows zero temperature behavior, while points are for thermal
  systems. Dashed line shows the bulk modulus $K_{A}$ for the same
  system at zero temperature. Solid black line in (b) shows $T^{0.66}$
  dependence while solid blue line shows $T^{0.5}$ dependence.}
\label{fig_plotGe}
\end{figure}

We calculate the pressure, bulk modulus and shear modulus for
two-dimensional random-bond networks with a connectivity of $z=3$ over
a range of reduced temperatures $T^* = k_{B}T/k_{\rm sp}\langle
l_0\rangle^2$ and bulk strains $\epsilon$. For these systems, the
critical connectivity is $z_{c}\sim 4$. Thus, our networks are
subisostatic and will be floppy at $T=0$ and $\epsilon=0$. Results for
the pressure are presented in Fig.~\ref{fig_plotPe}, for the bulk
modulus in Fig.~\ref{fig_plotKe} and for the shear modulus in
Fig.~\ref{fig_plotGe}. We first examine in detail the behavior of the
properties related to bulk deformation, i.e., the pressure and bulk
modulus, before examining the behavior of the shear modulus.

At zero temperature we find a strain $\epsilon_1$ (with corresponding
area $A_{1}$) at which the network just becomes rigid, indicated by
the solid black line in Figs.~\ref{fig_plotPe}(a) and
\ref{fig_plotKe}(a). Here the network exhibits a finite pressure and
bulk modulus above $\epsilon_1$, and zero pressure and bulk modulus
below, and we hence define this strain threshold as the rigidity
point. The pressure shows a linear dependence on area, increasing
continuously as $-P = c_{1}(A-A_{1})$ for $A\ge A_{1}$, where $c_{1}$
is a constant. Based on the definition of the bulk modulus given in
Eq.~(\ref{eq:bulk}) this means that $K_{A}=c_{1}A$ for $A\ge A_{1}$
and $K_{A}=0$ for $A<A_{1}$, i.e., a discontinuous increase in
$K_{A}$, corresponding to a first order transition from a floppy to a
rigid network at $\epsilon_1$. We note that the value of $\epsilon_1$
will differ for different network configurations, as there is no well
defined thermodynamic limit for random-bond networks due to the
average spring length growing with the system size. The response of the networks to bulk strain or thermal fluctuation doesn't differ  between different configurations with the same average conductivity. For the results
presented in Figs.~\ref{fig_plotPe}(a) and \ref{fig_plotKe}(a) a
network with $\epsilon_1 = 0.0356$ was used. The first-order nature of
the transition was present in all configurations studied.

When thermal fluctuations are present the network is rigid for all
bulk strains, as can be seen in Figs.~\ref{fig_plotPe}(a) and
\ref{fig_plotKe}(a) where different temperatures are represented by
the colored points. For small bulk strains ($\epsilon < \epsilon_1$)
the network is stabilized by the thermal fluctuations and exhibits an
increasingly large pressure and bulk modulus as the temperature is
increased. As $\epsilon_1$ is approached we observe a regime where the
pressure and bulk modulus for all temperatures start to join the zero
temperature line, with the low temperature results starting to join
the zero-temperature result sooner than the results for higher
temperatures. For bulk strains greater than $\epsilon_1$ there is a
mechanical regime, where tension is dominant over thermal fluctuations
and the resistance to deformation depends only on the spring
constant. However, we find that the pressure no longer increases
linearly with area as $\epsilon_1$ is approached, even at low
temperatures (see the inset of Fig.~\ref{fig_plotPe}(a)), resulting in
a continuous transition between the thermal-dominated regime and the
mechanical regime.

In Figs.~\ref{fig_plotPe}(b) and \ref{fig_plotKe}(b) we show the
temperature dependence of the internal pressure and bulk modulus in
the thermal, intermediate and mechanical regimes. Above $\epsilon_1$,
we find that they are both independent of temperature; in this
mechanical regime the network is completely stabilized by the spring
constant and its response to deformation is invariant to
temperature. At and below the rigidity point, the temperature
dependence becomes more complex. Below $\epsilon_1$ the pressure in
the network scales as $P\propto T^{\alpha}$. When the network is at
zero strain $\epsilon=0$ we find that $\alpha=1$, as expected in
analogy to entropic elasticity \cite{ref:deGennes}. However, as the
strain is increased we find $\alpha\lesssim1$, with an exponent that
decreases as the strain is increased, reaching $\alpha\sim0.66$ as the
critical strain is approached. We observe this dependence only at low
temperatures $T^{*}<10^{-5}$, with the pressure scaling linearly at
higher temperatures. This varying temperature dependence of the
pressure can be understood when we consider the behavior of pressure
in the initial linear response regime. That is, at low bulk strains we
find that the pressure scales linearly with area and at low
temperatures can be expressed as $-P = m(T)*(A-A_{0}) + c(T)$, where
$m(T)$ and $c(T)$ are constants for a given temperature $T$. It then
follows that the bulk modulus will scale as $K_A=m(T)A$. By fitting
this expression for the pressure to our simulation data we find that
$m(T)\propto T^{0.66}$ (for $T^{*}<10^{-5}$) and $c(T)\propto T$, as
shown in Fig.~\ref{fig_plotPe}(d). Hence, at low bulk strains ($A\sim
A_{0}$) $c(T)$ dominates and we find a linear temperature dependence,
while at higher bulk strains the system approaches a regime where
$m(T)*(A-A_{0})$ dominates over $c(T)$ and we hence observe a
$T^{0.66}$ dependence, with a mixed regime between the two. The bulk
modulus then scales with $T^{0.66}$ for all $\epsilon<\epsilon_{1}$ at
low $T^{*}$ and linearly at higher temperatures. On dimensional
grounds it follows that the pressure and bulk modulus must also have a
dependence on the spring constant and scale as $P,K_{A}\propto
T^{\alpha}k^{1-\alpha}_{\mathrm{sp}}$.

For bulk strains close to $\epsilon_1$, we find that $P$ scales with
the square root of temperature, $P\propto T^{0.5}$, for
$T^{*}<10^{-5}$ and linearly with temperature for $T^{*}>10^{-5}$, as
shown in Fig.~\ref{fig_plotPe}(b). In the $T^{0.5}$ regime the network
is again stabilized by both temperature and the spring constant, and
we find that $P$ scales as $T^{0.5}k^{0.5}_{\mathrm{sp}}$. We also
observe that networks below the rigidity point can enter this regime
as the temperature is increased. For these systems the pressure
initially shows a $T^{0.66}$ dependence before they then show a
$T^{0.5}$ dependence, indicating a regime that fans out from the
zero-temperature rigidity point. The bulk modulus, however, exhibits a
different behavior in this region, as for networks at $\epsilon_1$ we
find that $K_{A}$ is independent of temperature. For networks just
below this point we observe a rapid increase in the modulus with $T$,
before $K_{A}$ reaches the zero-temperature value.

As the area is increased for $\epsilon > \epsilon_1$, there is a clear
inflection point in the zero-temperature (and low temperature)
pressure, as can be seen in Fig.~\ref{fig_plotPe}(a). At this point
the pressure again increases linearly with area as $-P =
c_{2}(A-A_{2})$, where $c_{2}$ and $A_{2}$ are larger than $c_{1}$ and
$A_{1}$, respectively. This corresponds to a reorganization of the
network, as the nodes change positions to minimize the system
energy. This is illustrated in Fig.~\ref{fig_plotPe}(c), where we plot
the fraction of springs in the network that are \emph{activated}
(i.e., stretched or compressed such that $l\ne l_{0}$). At
$\epsilon_1$, we see the first springs become activated, followed by a
significant jump at a higher value of $\epsilon$. Furthermore, as the
area is increased beyond this point, we find several more
reorganizations, as can be seen by the kinks in
Fig.~\ref{fig_plotPe}(c) (there are also further kinks in the pressure
in Fig.~\ref{fig_plotPe}(a), although these are not visible on the log
scale used). This is present for all configurations and system sizes
studied, and in Fig.~\ref{fig_plotPe}(c) we present data from
additional configurations to illustrate this. This effect is also present when we take the spring
  constant of individual springs to be given by
  $k_{sp,i}=\mu/l_{0,i}$, where $\mu$ is a stretch modulus (see
  Fig.~\ref{fig_plotPe}(a)), such that very long springs
  will become progressively weaker and contribute less to the network
  response. This is likely due to the fact that it is neither the very long nor very short springs that dominate the system's response as the network is stretched beyond its rigidity point, which we confirm by examining the rest lengths of the activated springs in  Fig.~\ref{fig_plotPe}(c).
The effect that the reorganization of the network has on the bulk modulus can
be seen in \ref{fig_plotKe}(a), where we see that there is a second
distinct jump in $K_{A}$, corresponding to a first order transition as
the system rearranges, with further jumps present at higher areas,
although again, these are not visible on the log scale used.

We now examine the behavior of the linear shear modulus $G$, which we
obtained by shearing the networks at each bulk strain. For athermal
networks $G$ is zero at low bulk strains, as one would expect for a
floppy network before any of the springs become stretched. However,
the shear modulus remains zero beyond $\epsilon_{1}$, with the network
not resisting shear deformation until it reaches a bulk strain
$\epsilon=\epsilon_{2}$ ((see Fig.~\ref{fig_plotGe}(a)). This strain
corresponds to that at which we observed the second jump in the bulk
modulus as shown in Fig.~\ref{fig_plotGe}(a). Beyond this point the
shear modulus increases linearly with the area and the network becomes
rigid to shear deformation, indicating a continuous transition in
$G$. As for the pressure and bulk modulus, when thermal fluctuations
are present we find a non-zero shear modulus throughout, with thermal,
intermediate and mechanical regimes present, although here the
intermediate regime is found at $\epsilon_{2}$. The different regimes
can be seen in Fig.~\ref{fig_plotGe}, where we see $G$ remaining
constant with temperature above $\epsilon_{2}$ and $G$ scaling with
$T^{\alpha}$ at and below $\epsilon_{2}$, with $\alpha\sim0.66$ below
and $\alpha\sim0.5$ in the intermediate regime.

\begin{figure}[h!]
\includegraphics[height=0.4\textwidth,angle=270]{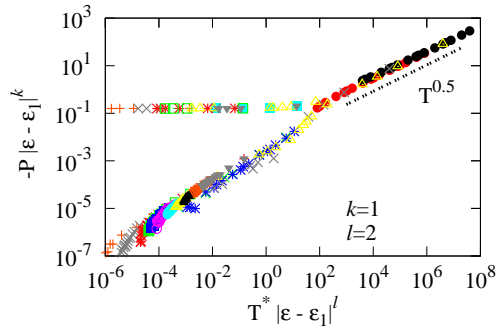}
\caption{(Color online) Scaling of the pressure $P$ using the form $P =
  |\epsilon - \epsilon_1|^{k}\mathcal{P} ( T |\epsilon -
  \epsilon_1|^{-l})$ where $k=1$ and $l=2$ are constants which give
  the best collapse of data. The two branches on the left hand side
  correspond to $\epsilon>\epsilon_1$ (upper branch) and
  $\epsilon<\epsilon_1$ (lower branch).}
\label{fig:scaleG}
\end{figure}

\begin{figure}[h!]
\includegraphics[height=0.4\textwidth,angle=270]{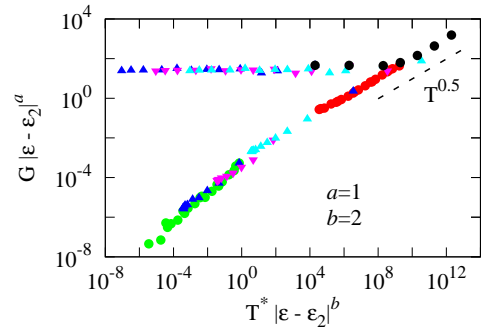}
\caption{(Color online) Scaling of the shear modulus $G$ using the form $G = |
  \epsilon - \epsilon_2|^{a} \mathcal{G} ( T |\epsilon -
  \epsilon_2|^{-b})$ where $a=1$ and $b=2$ are constants which give
  the best collapse of data. The two branches on the left hand side
  correspond to $\epsilon>\epsilon_2$ (upper branch) and
  $\epsilon<\epsilon_2$ (lower branch).}
\label{fig:scaleP}
\end{figure}

The temperature dependence of the different regimes of behavior for
the pressure and shear modulus can be captured by a crossover scaling
technique similar to that used for the conductivity of a random resistor
network \cite{ref:Straley}. This technique has been used previously to
describe the shear modulus for both athermal
\cite{ref:Wyart,ref:Chase_bend} and thermal systems
\cite{ref:PRL}. The scaling forms are given by
\begin{equation}
G = | \epsilon - \epsilon_2|^{a} \mathcal{G} ( T |\epsilon - \epsilon_2|^{-b}),
\label{scaleG}
\end{equation}
and 
\begin{equation} 
P =  |\epsilon - \epsilon_1|^{k}\mathcal{P} ( T |\epsilon - \epsilon_1|^{-l}),
\label{scaleP}
\end{equation}
where $a/b$ and $k/l$ are the exponents in the intermediate regime
for, respectively, the shear modulus and pressure. The best collapses
of the data are shown in Figs.~\ref{fig:scaleG} and ~\ref{fig:scaleP},
where we use the critical exponents $a,k=1$ and $b,l=2$. The two
collapses summarize the three regimes of network stability. The upper
left branches show the mechanical regimes, the lower left branch shows
the temperature dominated regime, where we find $T^{0.66}$ dependence
for the shear modulus and the varying $T$ dependencies for the
pressure, and the right branch shows the intermediate regime, where we
find a temperature dependence of $T^{0.5}$ for both $G$ and $P$.

\section{Discussion and implication}

The behavior of the sub-isostatic random-bond networks considered in
this paper is similar to the behavior found in Ref.~\cite{ref:PRL} for
lattice based networks. The observed sublinear scaling of the shear
modulus, $G\propto T^{\alpha}$, for networks below the critical bulk
strain was also found for lattice-based networks, albeit with
different exponents, with $\alpha\sim0.66$ for the random-bond
networks studied here and $\alpha\sim0.8$ for the triangular lattice
networks studied in Ref.~\cite{ref:PRL}. This indicates that, while
sublinear scaling is not confined to lattice models, the exponent does
depend on the topology of the network. In Ref.~\cite{ref:PRL}, it was
proposed that the scaling may be due to the internal pressure $P$,
which at $\epsilon=0$ scales linearly with temperature, leading to
$G\propto k_{sp}^{0.2}P^{0.8}$. This was in analogy to a study on
athermal networks with an internal stress $\sigma_{m}$ induced by
molecular motors, where $G \sim k_{\rm sp}^{0.2}\sigma_{m}^{0.8}$
below the isostatic point \cite{ref:Misha_motor}. However, as we find
that the pressure begins to scale sublinearly with temperature as the
bulk strain is increased from $\epsilon=0$ while the $G\propto
T^{\alpha}$ scaling remains, this proposed scaling would not be valid
as one moves away from the rest area of the network at
$\epsilon\ne0$. Indeed, the shear modulus shows the same temperature
dependence as the bulk modulus, which scales as $K_A=m(T)A$, where
$m(T)\sim T^{0.66}$ was obtained from the relation for the pressure
$-P = m(T)*(A-A_{0}) + c(T)$.

In addition to the similarities between the behavior found here for
sub-isostatic, sub-critical random-bond networks and sub-isostatic
lattice based networks, we also note the similarities between the
behavior of networks at the bulk strain threshold corresponding to the
rigidity point, and networks at the critical connectivity $z_{c}$. In
Ref.~\cite{ref:PRL} it was found that the shear modulus behaved as
$G\propto T^{0.5}$ at $z_{c}$ (at the critical connectivity the
critical strain is zero, $\epsilon_{c}=0$ \cite{ref:Misha_bulk}),
indicating that the stabilization of the network at the critical
strain is similar to that at $z_{c}$. We note that this is only true
of the shear modulus, as we find a constant bulk modulus for low
temperatures at $\epsilon_{1}$. A possible reason for the differences
in the observed temperature dependence between the two moduli would be
the nature of the zero-temperature transition from zero to finite
modulus, as the bulk modulus exhibits a first-order transition at
$\epsilon_{1}$ while the shear modulus exhibits a continuous
transition at $\epsilon_{2}$. We also note that the exponents found
for the crossover scaling anstaz in Eq.~(\ref{scaleG}), $a=1$ and $b=2$,
are more mean field-like than those found for the critical
connectivity case \cite{ref:PRL,ref:Chase_bend}.

Finally, the zero-temperature behavior of the random-bond networks
considered here differs greatly from that of lattice based networks,
exhibiting a non-continuous transition from a floppy to a rigid
network as the bulk strain is increased \cite{ref:Misha_bulk} and
exhibiting a regime where the system has a finite bulk strain but zero
shear modulus. However, despite these differences in the athermal
behavior, as previously mentioned the temperature dependence of the
thermal stiffening of the network does not change qualitatively
\cite{ref:PRL}.

\section{Conclusion}
In this paper we have studied the effects of thermal fluctuations on
the elastic response of random-bond networks at various bulk
strains. Our results show that, in agreement with previous studies,
there is a bulk strain threshold at zero temperature for which the
bulk modulus and pressure of a floppy network will become finite. We
find that the transition for the pressure is continuous while it is
discontinuous for the bulk modulus, jumping to a finite value at the
rigidity point. We have also found that random-bond networks can
exhibit further discontinuous transitions, as the networks rearrange
to minimize their energy. Unusually, the random-bond networks studied
here exhibit a regime where there is a finite bulk modulus but zero
shear modulus at zero temperature. In these systems, the bulk strain
threshold for a non-zero shear modulus is larger than that for a
non-zero bulk modulus, and the shear modulus transitions continuously
at its rigidity point.

When thermal fluctuations are present the network becomes stable for
all strains, and the pressure and bulk modulus transition continuously
between a thermally dominated regime and a mechanical regime at the
zero-temperature rigidity, while the shear modulus transitions
continuously at its own bulk strain threshold. In between these two
regimes, there exists a third, intermediate, regime where the pressure
and shear modulus depend on the square root of temperature (at their
respective strain thresholds) while the bulk modulus remains constant,
as the intermediate scaling occurs only at a continuous rigidity
transition. Perhaps most interestingly, we find that the shear and
bulk moduli exhibit an anomalous temperature scaling of $T^{\alpha}$
with $\alpha\sim0.66$ below the critical strain, where we would expect
to find normal entropic elasticity (linear temperature scaling
$\approx T$). This behavior is similar to that reported in
Ref.~\cite{ref:PRL}, where the shear modulus was found to scale as
$T^{0.8}$, indicating that floppy networks of various topologies can
exhibit anomalous temperature scaling.

\section{Acknowledgement}
This work was supported in part by FOM/NWO. We thank A.Licup, A. Sharma, M. Sheinman and C. Storm for many discussions.

\bibliography{main}

\end{document}